# Existence of Millisecond-order Stable States in Time-Varying Phase Synchronization Measure in EEG Signals

Wasifa Jamal, Saptarshi Das, and Koushik Maharatna, *Member, IEEE*

*Abstract*—In this paper, we have developed a new measure of understanding the temporal evolution of phase synchronization for EEG signals using cross-electrode information. From this measure it is found that there exists a small number of well-defined phase-synchronized states, each of which is stable for few milliseconds during the execution of a face perception task. We termed these quasi-stable states as *synchrostates*. We used *k*-means clustering algorithms to estimate the optimal number of synchrostates from 100 trials of EEG signals over 128 channels. Our results show that these synchrostates exist consistently in all the different trials. It is also found that from the onset of the stimulus, switching between these synchrostates results in well-behaved temporal sequence with repeatability which may be indicative of the dynamics of the cognitive process underlying that task. Therefore these synchrostates and their temporal switching sequences may be used as a new measure of the stability of phase synchrony and information exchange between different regions of a human brain.

## I. INTRODUCTION

Over the past decade, interest in the detection of brain dynamics has grown. Clusters of specialized neurons integrate to form a part of a big organization of segregated cortical areas which dynamically interact to steer the brain into specific cognitive states. This phenomenon of integration and co-ordination of interacting cortical regions is known as synchronization in time scale [1] and has been used as an effective tool for understanding the interactions between different brain regions during a cognitive task [2]. Quantitative Electroencephalography (QEEG) has been used over the years for estimating such synchronization between different brain regions by computing phase differences between the signals at different frequency bands acquired from the EEG channels distributed over the entire scalp [3].

Typically EEG frequencies are divided into different bands called δ (1-4Hz), θ (4-8Hz), α (8-12Hz), β (12-30Hz) and γ (30Hz and above). It is established in [4] that different mental processes and states yield responses in different frequency bands indicating towards the association of a particular frequency band(s) with a specific type of cognitive process. As an example, the β rhythm has been reported to exhibit links to cognitive processing, visual attention and perception related modulations [5, 6]. In the conventional QEEG process synchronization is computed over these individual bands and thereby estimating how each of these frequency bands are related at different brain regions during a cognitive process. However, to understand the transient dynamics of the information integration process in brain in a task-specific way it is necessary to estimate the evolution of phase relationships from the onset of a stimulus at different bands amongst different EEG electrodes and finally mapping them to the temporal evolution pattern of synchronization over the entire scalp. The fundamental problem of the conventional QEEG approach is that it does the phase estimation using classical Fourier Transform which does not preserve the temporal information of phase evolution and therefore gives an average measure of synchronization over a time window at each of the bands. In the recent years different measures of phase synchronization have been introduced [7] but are unable to preserve the temporal information about the evolution of synchronization. Therefore new measures are needed to estimate such transient dynamics of synchronization to understand the interactions between different brain regions subject to a specific task. In Quyen *et al.* [8] wavelet based time-frequency domain phase locking estimation of EEG signals is introduced. We have extended the concept and applied a clustering technique to obtain the underlying distinct phase synchronized patterns and also investigated their evolution and switching over time.

In this paper, we propose a new approach based on Continuous Wavelet Transform (CWT) for investigating the transient dynamics of phase synchronization in EEG signals. This allows effective investigation of the time varying instantaneous phase relationship leading to the detection of phase-locking periods between two signals. The proposed technique was applied to an EEG face-evoked dataset available online in the SPM website [9]. The data was preprocessed and CWT was computed for each EEG signal across all the channels. We explored the temporal evolution of the phase difference expressed as a function of frequency and time to find the periods of phase synchronization. The well-known *k*-means clustering algorithm was applied to cluster the phase difference vectors. A cost function optimization algorithm was run to find the optimal number of *k* for our clustering problem. Periods over which the phase difference has low variation are quantified as functional brain states where there is stable phase topography. Our investigation shows that there exist a small number of stable synchronization states of millisecond order, henceforth are termed as synchrostates. This is similar to the concept of microstates where stable potential distribution maps are clustered to form finite number of states [10]. We also found that the switching between these states follow a well-behaved stable temporal sequence from the onset of the stimulus. This measure can be informative in understanding transient neuronal dynamics and how it correlates to information

\* The work presented in this paper was supported by FP7 EU funded MICHELANGELO project, Grant Agreement # 288241.

W. Jamal, S. Das, K. Maharatna are with the School of Electronics and Computer Science, University of Southampton, Southampton SO17 1BJ, United Kingdom (e-mail: {wj4g08,sd2a11,km3}@ecs.soton.ac.uk).

processing of the brain and hence cognition. It also offers the possibility for its application to gather deeper insight into studies conducted on pathological populations.

## II. THEORETICAL BACKGROUND

### A. A CWT Based Measure for Studying Phase Synchrony in EEG signals

The concept of phase synchronization was first introduced by Huygens into the field of physics of coupled oscillators [11]. When in perfect phase synchronization the two phases of the signals are locked i.e. $\varphi_x(t) - \varphi_y(t) = $ constant. If the relative phase varies little in time, the two sources are considered to be synchronized. Phase synchrony analysis in EEG signals acquired over the scalp has been projected as an effective tool for understanding co-operative interactions between different regions in brain. However the conventional synchrony analysis in EEG does not preserve the temporal information and therefore offers average characteristics of phase synchrony only. Here, we propose a wavelet-based synchronization index which inherently preserve the temporal information and therefore gives an accurate picture of transient synchrony evolution from the onset of a stimulus.

Continuous complex Morlet wavelet transform when applied to a signal $x(t)$ yields a complex time series $W_x(a,t)$ which is a function of the wavelet scale $a$ and time $t$. The instantaneous phase of the signal at a particular time instant is given by the complex argument of the time series $W_x(a,t)$. Given two signals $x(t)$ and $y(t)$ their instantaneous phase difference at time $t$ is computed as (1) with $\varphi_x(a,t)$ and $\varphi_y(a,t)$ being arguments of the complex term of the $W_x(a,t)$ and $W_y(a,t)$ series respectively [12].

$$\Delta\varphi_{xy}(a,t) = \varphi_x(a,t) - \varphi_y(a,t) \qquad (1)$$

Although this technique of calculating instantaneous phase and formulating the relative phase between signals has been applied for the study of blood flow [12], here for the first time it is applied on EEG signals. After that the phase differences between the signals across different channels over time have been computed.

### B. Clustering for Finding Synchronized States in the Brain

The time varying relative phases (1) can be clustered into groups employing a certain class of pattern recognition algorithms. The $k$-means clustering is one such unsupervised learning pattern recognition technique. Given a dataset $X$, assuming that the number of underlying clusters is known, $k$-means is an iterative algorithm which minimizes the cost function

$$J(\theta,U) = \sum_{i=1}^{N} \sum_{j=1}^{m} u_{ij} \|x_i - \theta_j\|^2 \qquad (2)$$

where, $\theta = \begin{bmatrix} \theta_{T_1} & \cdots & \theta_{T_m} \end{bmatrix}^T$, $\|\cdot\|$ is the Euclidean distance $\theta_j$ is the cluster center and $u_{ij} = 1$ if $x_i$ lies closest to $\theta_j$; 0 otherwise [13]. Initially $k$ centroids are defined and the data vectors are designated to a class, depending on how near they are to the centroid. The $k$ centroids are recalculated from the clusters defined in the previous step. The data is reassigned to these new recalculated centroids. The algorithm iterates over this loop until the data vectors from $X$ form compact clusters and $J$ is minimized.

The optimization algorithm runs the $k$-means clustering $n$ times for each $m$ (number of clusters) in the range defined $[m_{\min}, m_{\max}]$ for the dataset $X$. For every $n$ runs the minimum value of $J_m$ is stored. If the plot of $J_m$ against $m$ indicates a significant "knee", it signifies the number of clusters that is likely to underlie the dataset $X$ [13].

## III. EXPERIMENTAL METHODS AND RESULTS

The analysis was done on the SPM multimodal face-evoked dataset [9]. Data was acquired from a single subject while the person was presented images of faces. The data was sampled at 2048 Hz and was recorded on 128 channels over several trials. For the purpose of this experiment we only used the first 100 trials. Epochs were created from -200ms pre-stimulus to 600ms after stimulus. All data was baseline corrected and an artifact rejection algorithm was applied to reject epochs over the 200μV threshold. Data was then band-pass filtered within 0.5-50 Hz. In order to compensate for the variability of our results, the whole dataset have been partitioned into three segments and then the experiment was run three times. We used data from trial 1-50, 51-100 and 1-100 for the first, second and third run respectively.

For each of the runs, the phase difference between all pairs of electrodes were computed by taking the argument of the complex Morlet wavelet transform of the signals on each channel and subtracting it from the other electrodes. For a more comprehensive analysis we converted the resultant complex series which is a function of scale and time to a function of frequency and time. This cross-electrode relative phase at a particular time instance is represented as a symmetric square matrix with zero diagonal elements as they represent the phase difference of an electrode to itself. These matrices were then averaged across the number of trials considered during that run. Observing these multi-channel phase data in sequence of intervals in the order of milliseconds indicate few patterns which are stable over finite number of time-frames. This is an interesting observation as it is similar to the concept of microstates in [14] where the authors observed stable potential distributions maps over millisecond order time segments. Similarly, we observe that the phase maps remain stable for certain time periods and rapidly change to new configurations that are also stable over certain time frames. We define these stable maps as synchrostates during which a subset of the brain regions is synchronized and it steers itself to perform certain functional tasks – in this case, a face perception task.

In order to estimate these states optimally, we perform $k$-means clustering for each phase matrix, to classify similar states into a single class following the method described in Section II B. Hard clustering algorithm is applied with the assumption that the brain can be at only one state at a given time. We estimated the synchrostates for each EEG bands (θ, α, β, γ) by taking the average response over all the frequencies of each band. Fig. 1 shows the results from all the three runs of the optimization routine for optimally clustering the synchrostates in the β band for illustrative purpose. In this case over all the trials the *k*-means clustering results into three unique states as there exists a "knee" in the cost function at $k = 3$.

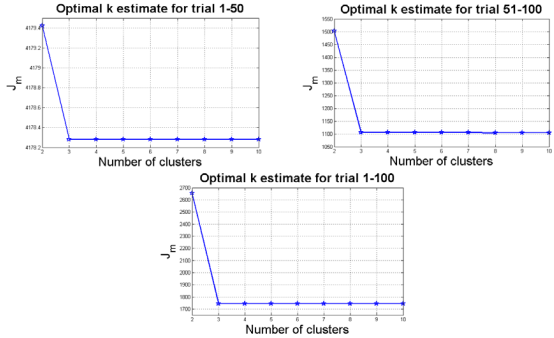

Figure 1. Optimum *k* values for different group of EEG trials.

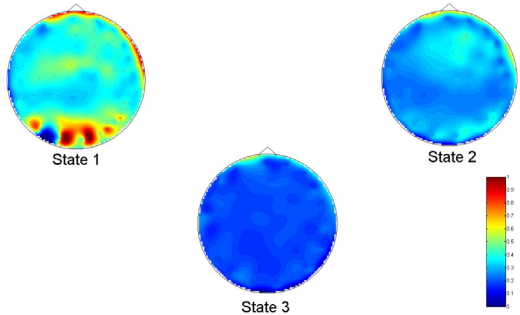

Figure 2. Clustered synchrostates for trials 1-50.

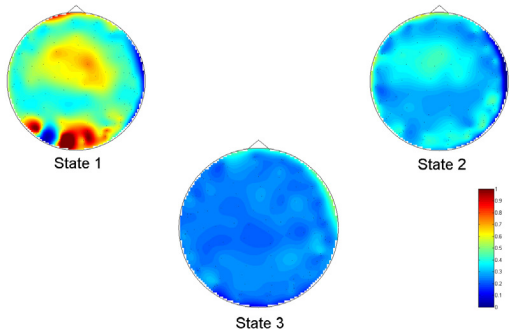

Figure 3. Clustered synchrostates for trials 51-100.

To get a better idea of the topographical distribution of these three states, we plotted the normalized average phase difference of each state over a head model illustrated in Figs. 2-4. It is also interesting to note that the topographical maps of synchrostates are consistent and are almost unique in our simulations. In order to estimate the time course of each synchrostate, we plotted their changes over 400 samples (approximately 195 ms) after the onset of the stimulus. The time-course plots for the three runs are shown in Figs. 5-7 for the runs on the dataset with trials 1-50, 51-100 and 1-100 respectively. As can be seen from Figs. 5–7, the time course of switching between these three states shows consistent sequence pattern with similar time period.

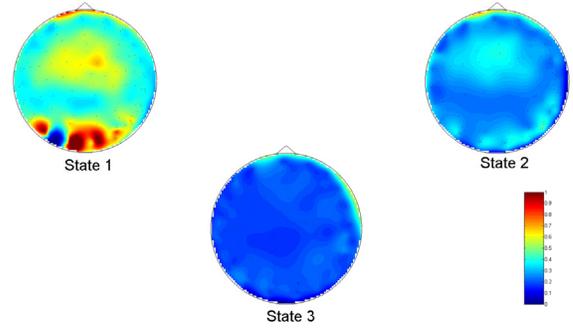

Figure 4. Clustered synchrostates for trials 1-100.

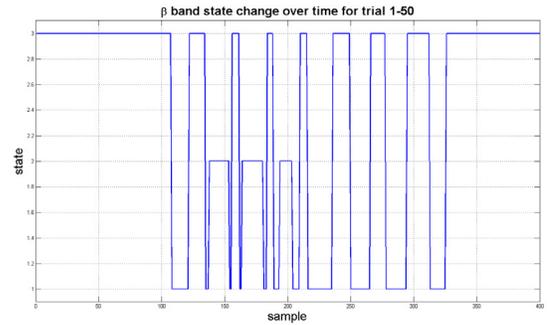

Figure 5. β band temporal evolution of states for trials 1-50.

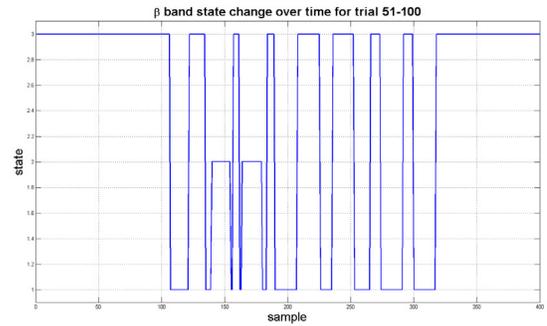

Figure 6. β band temporal evolution of states for trials 51-100.

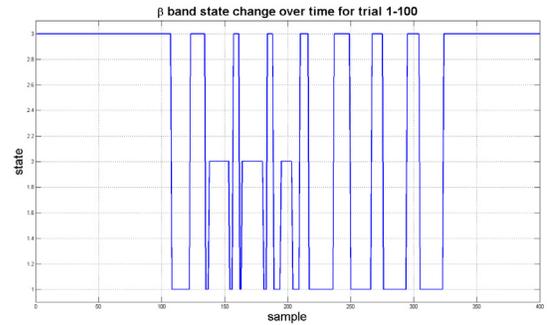

Figure 7. β band temporal evolution of states for trials 1-100.

To explore the repeatability of the synchrostates for the present task, we computed the number of times each of them occurs in the β band as shown in Table I confirming that the number of occurrence of each of the synchrostates is consistent over separate trials with small standard deviation. The little variation observed could be attributable to the fact that even during a focused task, there could be multiple cognitive processes that run in the background. These are not directly related to that specific task but may influence the phase relationship between different brain regions in an indirect way.

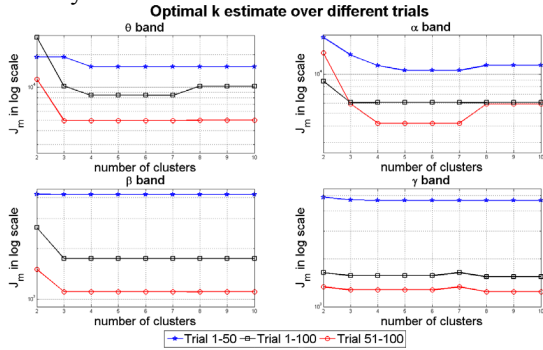

Figure 8. Cost functions for clusterings in different EEG bands with increasing $k$.

TABLE I.   TIME OF OCCURRENCE OF THREE STATES FOR BETA BAND

| Time of Occurrence (number of samples) | State 1 | State 2 | State 3 |
|---|---|---|---|
| trial 1-50 | 101 | 43 | 256 |
| trial 51-100 | 105 | 31 | 264 |
| trial 1-100 | 113 | 42 | 245 |
| Mean | 106.33 | 38.67 | 255 |
| standard deviation | 6.11 | 6.66 | 9.54 |

We applied the same technique for estimating the synchrostates in the θ, α and γ bands and the results are shown in Fig. 8. We found that in θ and α band the optimal number of synchrostates varies between separate trial runs but within a small range of numbers (approximately 3-5) whereas, for γ band the optimal number of synchrostates is obtained at $k = 3$ consistently. In terms of time-course of the switching sequence, for each of the bands, its characteristic pattern may be similar over all the trials.

## IV. DISCUSSIONS AND CONCLUSIONS

From our simulation results it is evident that there exist a small finite number of states in different EEG frequency bands that are indicative to millisecond order stable phase-locking topography. Physically it means that during an information processing task – in this case, a face perception task – a well-defined information exchange process is initiated between different regions of brain and the state transition characteristics resulted indicates the dynamics of such a process. A very small variation in the number of optimal synchrostates is observed in the θ and α band between the trials (typically ranging $k = 3 - 5$). This could be attributable to the influence of random background cognitive processes which are not directly related to that specific task but may still be present during the its execution. The inter-synchrostate transition occurs abruptly which is also similar to the transitional nature of microstates [14]. Assuming that each task can be broken down into a sequence of subtasks, this may mean that the time duration of each synchrostate in the time-course sequence is indicative to the processing time required for the underlying brain circuitry for processing a subtask. Therefore this could be an effective tool for quantitatively characterizing information processing ability of brain in neurophysiological disorders like Autism where information integration and processing speed are the biggest issues.

In conclusion, although our results show existence of small finite numbers of synchrostates exhibiting well-behaved switching sequences amongst themselves, more rigorous neurophysiological investigations are needed in task-specific way in different pathological populations to understand the full implications and applicability of the synchrostates. Two fundamental questions are whether the synchrostates are task-specific or person specific. Since this is the first work of its kind, here we present the basic observations obtained from face perception task in person-specific way. In our future works we will investigate the above mentioned problems in more detail.